# Efficient management of IT Infrastructure implementation and support at enterprise level

Bhargav. Balakrishnan

Sutherland Global Services

# 383, Velachery Tambaram Main Road

Vijayanagaram, Velachery

Chennai - 600 042

Mobile no – 9840658679

bhargavbalakrishnan@gmail.com

Abstract: - This paper deals with how to manage effectively in the design, implementation and support of an IT infrastructure at an enterprise level. This particular management is lacking in today's IT infrastructure scenario. Just implementation is not sufficient for an NON-IT industry, they need a proper support in the infrastructure like documentation, support work flow, ticketing systems (used for IT related issue either hardware or software) etc... Many organizations spend a lot of money for this support and they expect a lot from the provider. Many providers sign in the SLA that they will provide them with an excellent support, but 80-90% it doesn't happen. Many times they don't meet the expectations of their client. So how to make these expectations being met 100% for the client? That is what is going to be discussed in this paper with respect to ITIL framework and other technical terminologies.

# INTRODUCTION

In today's world every IT and non-IT organization's depends on IT for their infrastructure. The important thing will be the documentation of the entire infrastructure both hardware and software. The documentation should include the entire work flow of its implementation i.e. floor planning, network cabling, switch rack diagram (Support agreement of networking devices), servers, storage etc.., so each procurement will have their own warranty period. This documentation is called as an Asset Management. This asset management is not properly managed in many enterprise organizations and licenses are being misused by the administrator or employee of an organization. Let us take an example of Windows XP OS. Normally an organization at an enterprise level will buy not less 1000 licenses. In that case if someone misuse, it can be tracked easily through Microsoft if needed, but the management have the faith on IT support team that they clearly maintain the records. When it comes to the IT audit, those one or two licenses misused are not cared as there are n numbers of software's used at enterprise level and due to lack of time they normally skip this issue. This software's should be automatically deployed from Microsoft, Infosys etc.., for their product that they deal with their client. The organization should order the licenses online and get it deployed automatically from the respective dealers. Why they give this to IT-support with lot of expectations? The main thing is time, which is what matters a lot in today's world. The time you take to install an application is going to contribute 0.0000001% of an organization's profit indirectly. So IT-Support is very important for an organization once after the implementation is done. So here in this paper we are going to see how we are going to effectively balance the management of IT-infrastructure implementation and support.

# MANAGEMENT OF IT INFRASTRUCTURE IMPLEMENTATION

When an IT infrastructure is designed, a proper case study needs to be done on the basic requirements like floor diagram, network ports requirement (data & voice), network devices required, servers, storage and other software's required needed to be documented. This should have proper approvals both for procurement and dispatch of the products.

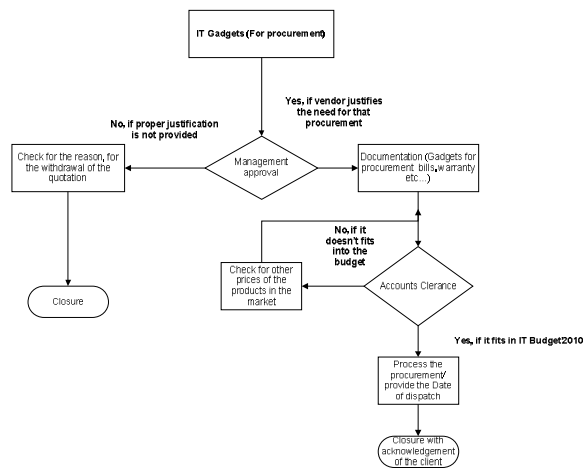

Fig (1) Management and accounts approval and clearance process

In the above diagram, it shows a basic work flow for a procurement of "IT-Gadgets". Similarly the procurement needs to be done for the entire requirement. This documentation and approval process will be very helpful during the audit. This will enhance the clarity for the management regarding the asset that they have regarding the IT. In the diagram it shows the approval from accounts as they are going to provide the finance for the procurement of IT gadgets. These things needs to known to the accounts and management that the following gadgets are brought with their approval and it has been bought within the budget of that particular year. **Why ITIL is useful here?** ITIL is a framework as we all know and it is been universally accepted. Here ITIL plays a vital role in providing a proper service strategy, service design and service transitions. As these three stages are to be properly placed to go ahead with service operations and continual service improvement which comes in support.

1) Strategy
2) Design         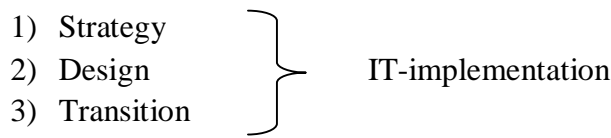 IT-implementation
3) Transition

1) Operation                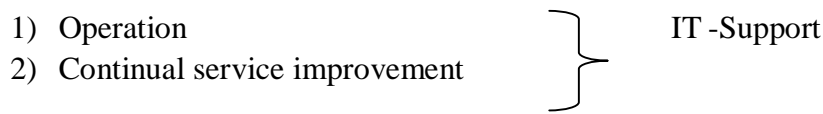 IT-Support
2) Continual service improvement

*Usage of ITIL in implementation*

How the first three of ITIL is related to the last two?

Let us discuss this with an example of an organization which going to start with 40 employees. At first the management, finance and IT manager or third party vendor need to make an initial documentation of what is required i.e. service strategy. This needs to be refined based upon the quotations provided by two or three vendors. The selection of vendors should be specific that is nothing but the vendor should be an authorized dealer of the products which you are going to procure from them. The agreement that is signed should be reviewed properly by the top management before proceeding with the procurement process.

*Process Flow*

Case study of the requirement -> documentation for management review ->filtering of based on quotation and authorization of the products that are going to be procured -> final documentation for the top management review-> once approved for both procurement and support (within warranty and after warranty i.e. through SLA) -> Closure of LOP

The support provided by the respective vendor of the product should be analysed on half yearly basis by the IT-support team and has to be forwarded to the management for their notice. Once this is approved by them, then only the renewal for the support needs to be done. This needs to be web based application for the IT-support vertical. The updation related to all the vertical needs to be done by the IT-support which will reduce the time to process them manually. All the bills should be made web based so that it can be made a centralized process for both accounts and management.

*Process Flow*

Support provided by X dealer for six months -> IT-support review – Satisfied (in the scale of 5) (4) -> automatically sent to management and account staff-> review with the vendor why 100% support was not met-> reason of both IT and vendor are justified-> closure for renewal (Decision by the management)

Decision will be based on if the vendor has provided 100% capacity and availability management for the client even then the required target is not met. The other reason will be providing less capacity and availability management and resolving the issues that can be only upto to their level i.e. capacity. In the second reason the capacity i.e. no of technical support executives was less and they were about to give 100% efficiency for their capacity, so either ways the vendor has not met the client target. This is the place where the organizations have not gone into in depth analysis. There had been lot of loss for the organization directly and indirectly in their profits and targets that they have set for that particular year at least by 1-2% difference. This can be properly framed with a lot of review at the service strategy stage itself. The vendor has provided a proper documentation that is whatever is required by a new organization is quoted with proper support both during warranty and after, along with the resource he would provide for the support. Example TCS providing support for Ashok Leyland. His major concern will his core application, email and website which needs be active throughout the year without any issues as it is going to grow his business. Whenever the applications go down or any issues which are major in that circumstances he will expect the proper resource from the support organization. These kinds of mistake in many situations have lead to closure of their contract along with loss for the organization. These things need to be properly framed and approved by management.

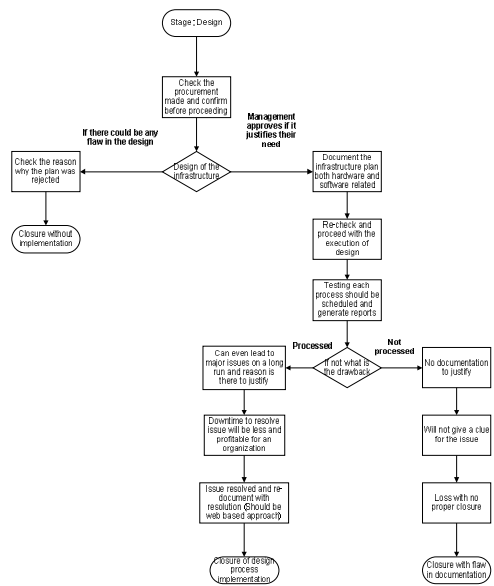

Fig (2) Importance of a design process

After this process will be design of the infrastructure, which will be done based upon the procurement done at the initial stage. The management, IT staffs and vendor needs to make a clear analysis that how the design of the infrastructure needs to be done. The floor diagram, network diagram etc...based upon which the procurement is done. The design should involved the approvals from the management level before implementing anything like DB, telephone exchange box, trunking of cables as in present world there are lot rules and norms that needs to be followed while setting up an infrastructure and each component has it's own space. The DB when it is set, the total load required for the office will be calculated separately and for server room it will be dedicated one. The power of the floor or building should not be given to server room as it is a room that requires lot of safety. It is called as the heart of an organization. This is where the design should be done properly, it involves an IT in both electrical and cabling design for an organization. The IT staffs should fix the servers, storage and switches on to the rack, then calculate the load it is showing on an average. Then the load should be allotted double of it. This plan has to be sent to management and justify why they need this load and how they got this value. These things need to properly document with a final copy.

Then comes the network cabling and it also varies based upon the management requirement also. Sometime for the conference room they will go IP projector, video conference and ports that are placed on the conference table for their clients to use. All these things need to be properly fixed before going ahead for final execution of network cabling. The data and voice ports need to be properly documented as in long run this will be easy to trace. Whenever there is damage in a port it can be easily troubleshooted by the support team. The management should refine all these requirements, validate it and approve. Then the IT team can go ahead and finish the design. The IT team should check the load of the server rooms on a regular basis that is during peak and non-peak hours. According the DB usage can be seen on regular basis. These documentations should be given to the management to justify them that whatever is done at the stage of strategy and design has been properly utilized. Based upon the trial that is done by IT team say for about 3 to 4 months period the change required in the design, installation and documentation can be provided to the management. Then these changes need to be properly documented with proper justification.

**CHANGE MANAGEMENT**

Then final stage of implementation is the change which depends on the periodic report generation of the infrastructure which can be both hardware and software. The change depends on the requirement but there is a proper approach behind it which are mentioned below

1. Downtime (which can affect the business by stopping the service that needs up gradation)
2. Risk involved (information in perspective in loss and confidentiality)
3. Documentation updation and test run before executing it on live.

So the change should be bought as and when we require. It needs to be approved by the management by seeing the above criteria's and other factor involved behind it. They would see what would be investment on return if the change is brought about. What are the other solutions for this application? Is there any other application for performing this task? How it gets integrated and who is the provider of this? Can this be provided by the same developer say TCS, Wipro? What will be the cost involved? These are some of the points that needs to documented with proper justification and then sent across to management to go ahead with the change. Once that is done with the task needs to be scheduled and intimated to the staffs at least 72 hours before and documentation should be sent with a screenshot that where the change is brought about if it an software or a hardware. This documentation should be updated on a whole with all the changes every quater for an enterprise organization which gives the proficiency of an organization in maintaining the IT records. This will what is going to help in the support stage which is going to discussed in the next and there we will see how these stages build a foundation for the support vertical.

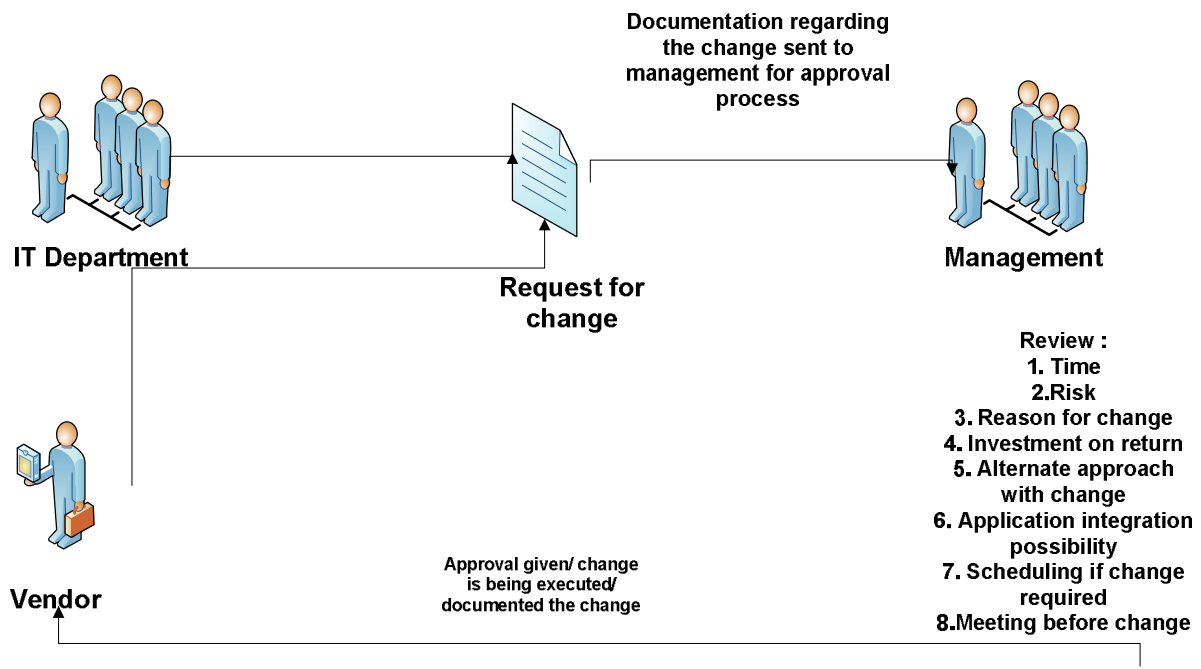

Fig (3) Process involved in transition

*Change Process Flow*

The change advisory board is going to decide the change that is going to be made in a particular software or hardware. The change management team should check the following criteria's like risk involved, downtime, and bugs if it occurs what's the alternate application etc... These things needs to needs to documents and approved by the head of the change advisory department hear to the board of directors. The documentation should be precise that is it tell what is required, all criteria's, need for change, alternate solution's, investment on return, justification for change. Once approved should direct a copy to the vendor, change advisory and finance department. Once it goes to the finance department they need to check on the budget and the emergency of the change. Based upon the time frame needs to be set and the change needs to be brought about.

Change is like a bridge between implementation and operation's so that is going to play a vital role. Documentation in this stage is going to tell about the changes that were made earlier on any particular application. The strategy, design is the foundation of an organization's start but transition is going to stay with the organization forever. Every decision is going to be carried out based upon this stage. The things are looked by swapping the information's that are stored in the documentation of this stage. How to make this stage a perfect one? There are people from all the departments in this stage, but there should be a team which should consolidate the opinions of all the member, make a proper a documentation on the entire transaction of each change right from approver, vendor, the area of change, the bugs that were involved in change, the solution for that bug, the importance of that bug, the downtime caused due to this bug, the department affected by this bug etc.. All needs to be documented and sent as a alert message to the team members of that department. When the change is being carried out, the system administrator should check that alternate application is working properly and that day's work is not been disturbed by this change of this application. Immediately after the change, the test run should be done by using the dummy input value and then with the approval for release the application should be moved on to live. This approval should be within 2-3 hours. Once done the users should be intimated about the location where the change that has been made.

Change management should be carefully handled by the management as the changes made should be one time; it should be tested in that way. There should be no bug that is 99% that shows the perfection and proficiency of the vendor in providing a change at the time of need. That is what the Non IT organization expects from an IT organization. For this they spend in thousands and the expectation will be error free. Once that is met then the 75% of the SLA agreement is met. Changes are the one that is mainly seen by any organization and how well it is documented and supported in the time of need. These documentations are the one that is going to provide a proper resolution for any issue and also reduces the time spent in tracing the location of the issue. If the error occurs, first will check is there an error in the changes that is being then proceed in depth into the application and fix the issue. This is what that needs to looked into while bringing about the change.

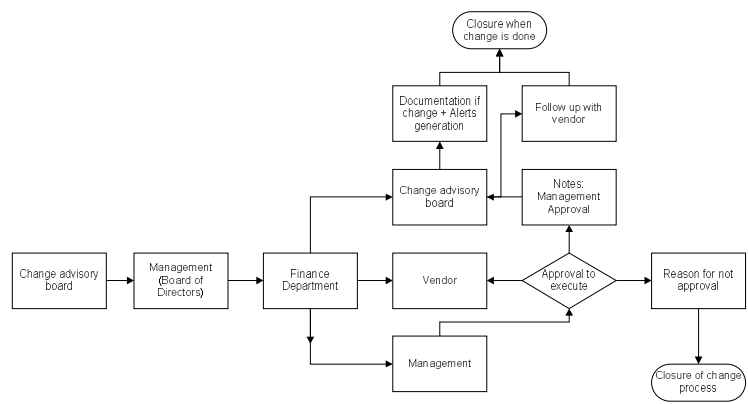

Fig (4) Process Flow of change management

Sample output of the application (Web Based)

Fig (5) Home Page for strategy

Fig (6) Procurement (For Management)

Fig (7) Approval Process (Management approval)

Fig (8) Approval Status

Fig (6) Requirement detail (.XLS file)

**Design Phase**

Fig (9) Notification of approval to vendor

Fig (10) Vendor notification email

Fig (11) Approval for design process

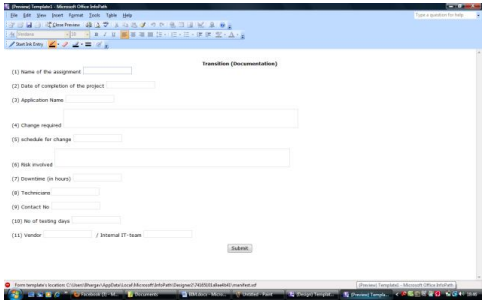
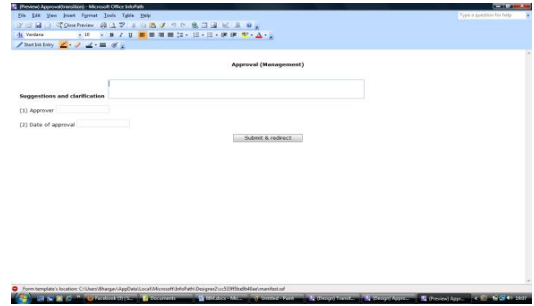

Fig (12) Documentation for transition

Fig (13) Approval Transition

The above shown sample screenshots describes the important process flows involved in the implementation stage of an IT-infrastructure. Once it is done with then it moves to the next stage of IT-support.

**IT-OPERATION**

*Support*

Support is purely based how good an implementation and documentation are done. When the documentation is done as per the flow then it makes it clear for a system admin to take over the task within a week's time. Let me provide you with an example of a documentation of a server.

Server:

(1) Name – HP Proliant DL 380 G5 ( 8 SAS HDD with each 146GB capacity)
(2) Configuration – Xeon processor/ 2.73 GHz/ 8GB RAM
(3) Operating System – Windows 2003 (R2) server OS
(4) Application – ADS( User accounts, group policies), DNS, DHCP, Print Server
(5) Anti Virus – McAfee 8.5
(6) Policies Applied  - i – Prevent accessing control panel ii – Prevent modifying date and time iii) Software Deployment (Ms Office 2003, McAfee 8.5, Win Zip, Adobe, Internet explorer 8.0)
(7) DHCP Range set – 192.160.2.10 – 192.160.2.225
(8) Print Server – HP Colour LaserJet 4600n, HP LaserJet 4100n

The above mentioned is the procedure how a documentation should be made. This should be precise and will help the support team to maintain the updation properly in this order. Every day the logs needs to be generated for every task like security, web access report, application log, NT authentication logs etc... Once this is done with then the support process will be smooth and the continual service improvement will be there for the organization as the vendor is meeting the target that he has agreed to deliver to the client without any flaw. There will be a fault tolerant but in the range of max 0.5- 1%.

Support is what going to decide the continual improvement management and further transition is going to come. Because whenever the support documentation that are prepared should be checked and based upon which the change needs to be carried. If a user account in Active directory services, the System Administrator should check what application is required to that user, what permission needs be set that is the group policies, access restrictions based upon which the budget is normally calculated for IT for a year. There should be forecast report that needs to be generated by the HR and that should be map with the account's budget. These put together should ask for the hardware spec that is needed from the vendor. Then this should go for approval and get that fixed. What is the role of support here? Support has to monitor even this requirement as he is seeing the live operation and the growth in the requirement. The asset is being managed by the administrator and asset management team together. Once the reports are generated on a quarterly basis, it will be easy for an organization to take any decision on IT-budget. That is one of the major role of support.

Support operation should follow the flow of process that is Issue -> Analyze the issue -> provide appropriate solution -> Case creation + documentation -> closure if resolved nor escalate to Level 2 or 3 based on the emergency of the issue. How the documentation of this should be done. The documentation should be clear and precise that is as follows

1. What is the issue?
2. Username and IT tag no ( Asset management # which is also the computer name)
3. Root cause of the issue (If identified from the previous documentation)
4. Risk Level (Mark the emergency of the issue, tag them accordingly)
5. Resolution (If resolved)
6. Closure/approval (depends on the type of issue – Solved / procurement of any application from the vendor)

These are some of the notable documentation contents that are made by the support. The arrangements of all these issues put together are going to reduce the downtime while resolving the any issues. If a issue is not getting resolved by a specific resolution, the support tries some other approach and if it get resolved, then that note has to follow the previous one. This is how the continual service improvement and requirements for more services will be put forward by the client to the vendor say TCS, Wipro. If this is getting failed and every time if the issue is fixed as a patch then it is no use in providing further service rather switch this support operation to some other better organization. A good example will be Oracle, many organizations support oracle implementation. If a x organization implements with a lot of bugs and above that there is lot of failure in support then it is matter of approval to make a change to another vendor Y who also deals with oracle but here what matters is the what was investment on return. They will now make a lot of studies and analysis on the change that is going to be taken as the existing X has not provided a satisfactory service. Once that is cleared then they will step forward for a change. A vendor decision by an organization should be clear and with proper approvals. What are the services they provide, other services can be got from them if the service provided by them is satisfactory and has met the level of agreement need to analyse before taking a decision. Once that is done, then the organization can be sure that the vendor will provide those best of service both in support and implementation. How the support process flows.

Support team -> IT management (client) -> Issue came (quarterly basis) -> resolved /unresolved issues -> unresolved (Reason for not resolving) -> % of resolved issues + Downtime + fixed permanently or temporary.

*Decision Flow*

In the above decision flow it shows how a vendor contact IT team of client and produce the documentation of the issues handled and resolution %. That needs to be mapped with the SLA agreement and needs to be approved by the management. This need be carried out on a quarterly basis and then consolidated at the end of the year. This will show how much of care that the vendor has taken in supporting the client. Based upon this report only the next year contract will be renewed by the client. This SLA period will vary will different organization. But the resolution percentage and downtime are the main criteria's that are taken into account. Every calculation right from the payment, agreement renewal, and other services requirements all depends on these two criteria's. The documentation should be done and needs to be forwarded to the IT team of the client. When the issue is not resolved based on the emergency of the issue and the risk involved in that the escalation process needs to be carried. A good example will be Oracle again, if the issue comes in an oracle application which is implemented by the oracle partner say Wipro, TCS or HCL, if they are not able to fix the issue which is highly important say approval button of CEO, then the time frame will be less, and it should not search in Google or within the team in the vendor. It should directly be escalated to expert level support in Oracle and get it fixed within ½ to 1 hour. That is how the flow of this process should be. Even if the technical person in vendor knows, he should immediately attend and fix it without any issue. If he fixes the issue and other option doesn't work then it is nothing but screwing an application. Each management level member have their own importance, as say with the approval of CEO the end of the day report needs to be generated, then it should be handled only by the expert level team even though the issue is known to the technical level team of the vendor. These are some of thing that will add positive signs for the vendor. This is what going to count in % of resolution + downtime. There should be a column for critical issue handled and resolved.

This will show how much of care a client has taken in fixing the issue for their client. These are some of things the client will keep in his mind and renew the contract. When there are changes during the support (IT-operation), there also the downtime matters, it will involve the same criteria's that I have mentioned in change management. Only difference is the time factor. As there the stage was in implementation, so the technical team took time to check and modify. Here mostly it comes in emergent requirement that needs to changed or deleted within 24 – 48 hours. Operations will not stop; we need to work on that working application that is the real challenge us IT professionals have. As a server even though it works on redundant policy, when a change is made they will be a disturbance when the synchronization happens from the main server. That's the reason why we may do most of the operations in night, but certain operations needs to be carried out with a lot of risk that is net banking as it is accessed 24*7 by people from all around the world. If a add feature to a website on one location, it needs to get replicated on the other location where there is redundant copy of this application. How to frame a process flow for this situation can be seen below.

IT Team -> changing/adding a feature (Net banking) -> Check where the other servers are present (From Documentation) -> Make a note of risk, downtime -> Send an alert message to all users (automated message) -> Then proceed with the change after providing the alternate website _-> once done send an automated message to all users. -> send that within 15 minutes downtime

*Process Flow*

Here you can see how an IT team conveys the message to their valued customer when there is an outage of a service. These particular automated messages are not seen in today's organization. When there is an outage, it can be seen when we access their website. It will show a message "access our website after 24 hours". This is one way to convey the outage, but this won't help the users to know immediately. That is where an automated message needs to be triggered directly to the users contact number.

These are some of things are seen in depth while it comes into contract renewal. These extra features have provided a splendid support services for the valued customer of the client that will be taken into account during the process of renewal. So the support for issues, downtime and feature used for notification of outage are the main criteria's that will make the client happy as he will be valued in turn by the customers. Customer accepts the best service and satisfaction in using their services. Having features are not important but using it at right time is important. When the vendor is able to monitor all these logs and produce them at the end of the year with all documentation, that is what will show the entire support provided by the vendor is satisfactory. A survey should also be done among the employee to the service of the vendor from 1-5.

When the support is provided, the tools used to resolve the issue should be like logmein, VNC viewer, and team viewer for remote support. Database storage is important; a parallel copy of it needs to be stored at the vendor side after getting the confidentiality agreement signed by the organization. This is how the support at the operation level should work. Each and every task involved in support should be fixed with the time frame and ticketing systems that each issue should be tagged with a number. This should be tagged with different tagging code like application issue aplXXXXX (x is the number) / hardware issue hrdXXXXX. This will classify the issue and easy to refer when the issue comes back again. Let us now see the flowchart of the support (IT –operation) at enterprise level.

In the diagram below we can the difference when the issue comes to a single user how the process goes and when it is for a department how it is routed through. That is based on the importance of that application for the respective user. Microsoft office, WinZip, adobe are some of the most commonly used applications If the problem comes of one user, it can be management; the % of service affected will be less when compared to entire department,

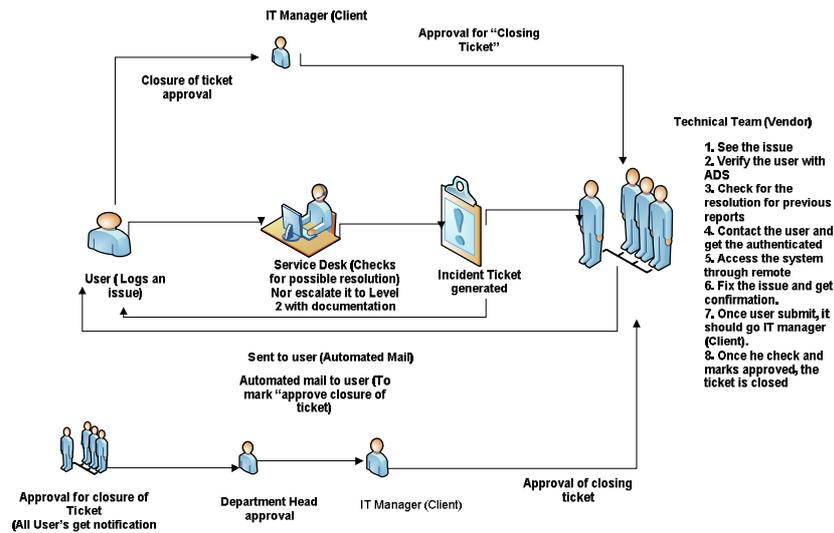

Fig (14) Approval process in Support (IT-operation)

Let us know see how this methodology is going to be effective both in implementation and support. The advantage are mentioned below

1. In the methodology, the entire process is mapped with ITIL, which is universally accepted framework that an organization should have. When an implementation is done, the most important thing will be "Which Framework" have you followed to establish this infrastructure.
2. Each stage in this methodology is wrapped up with a lot of approvals this is only to make each and every documentation that is being transformed into reality are properly conveyed and accepted by the board of directors.
3. There are many places where I should have mentioned Time as a major constraint is only because in this today's scenario everyone want the task to be performed within the timeframe that is accepted for in the SLA, let it be strategy, design, transition or operation, the time matters a lot for the final finish that is continual service improvement. Only when the constraints are met, the client will be happy to give the vendor an opportunity to provide other service which they have, which is in turn required for the client.
4. This way of approach will help the vendor to attain success in both implementation and support. As there are lot of practises which not been inculcated in reality are brought about in this paper like notification over message to the user that there is an outage, this can be seen in very few banking sectors. When a person goes to bank, the bank staff, today our application is not working. How frustrating will be to a customer, but for a staff he will get his one day salary. Like this many things are been modified and have made it as unified process.
5. Each process in this methodology is closely tied to each other, without properly closing the initial stage; the next stage can't be started. Each stage has approvals and documentation. Those documents are refined again by the management in order to have proper implementation and support phase.

6. Finally by following this approach will not bring success but a proper way of documenting things which are carried out in each phase, the most preference will be given to operations and transition as they are going to work parallel in the live environment. They are the pillar of every success of the client and vendor relationship.

When this type of methodology is adopted, a lot of flaws can be avoided in the implementation and support. Once the support is attained to 99%, there comes the talk on continual service improvement.

**CONTINUAL SERVICE IMPROVEMENT**

The term continual service improvement itself say's improve continuously while climbing up the ladder of success. When an organization is satisfied with the reports generated by the vendor, definitely the faith on them will become higher and higher. Then they will have a word with their board of director on how to modify or upgrade the versions of application, other better technologies and their advantage over the previous ones. Along with that they will also see the investment of return. If the organization is going for VOIP they will see from an ordinary phone, what will be the added advantage in terms of fault tolerance, throughput, compatibility, features like call forwarding, IVR system, license charge, hard or soft phone, manufacturer, support in terms of local and global which will be documented before going ahead for approvals, design, requirements, and implementation. All these put together will be calculated in the continual service improvement. At every stage, an organization would like to make their infrastructure better in order to reach the other market faster. That is the business tactic of spreading their product, for which they would have rely on a vendor. If the vendor provides the best support then providing the service improvement plan to him will their aim.

In the continual improvement phase also the risk, time factors are calculated. This is also a type of transition where a new technology, options are added to the existing infrastructure. Once this is done with the entire flow of IMPLEMENTATION and SUPPORT will be successful. This is the main goal of this paper. The continual service improvement itself tell how the service has been provide to the client, where they satisfied with our service, and have we met the service level agreement set by the client. The entire paper has showcased the process and their relationship with each other. Each process has to be properly studied while documenting and while transforming theoretical information in reality. As transformation of information into reality only shows how perfect we have implemented the infrastructure, changes that are needed and the key area of monitoring during support. Once these are properly set the success is being met and the result can be seen in the growth of the organization (client). Our aim should be providing the service to our client and finish it to their satisfaction. The same way the support has to be rendered to the level of satisfaction and with appreciations that support provided by the X vendor was splendid and now we will give them a chance for them to work in our projects. To attain this, a methodology like this will be standing as a pillar. This was the main focus of this paper which includes both implementation and support packed up by ITIL framework.

## CONCLUSION

The main focus of this paper was how to provide an effective way of IT implementation and support at an enterprise level, where these phases play a vital role. It needs to be properly structured with the frameworks like ITIL, Cobit etc... Once the implementation, the testing's needs to be carried out first at the first three stage as a whole even though at each stage we have refined, got approvals and collaborated the requirements to move into the live environment from the testing phase. At the transition it is quite important as the changes are to be document. Once that phase is done, support has to a proper process flow where time frame should be an important constraint. Once these are attained as per the process flow mentioned in the topics above, continual service improvement will automatically be given to us as the SLA is met as it was mentioned in the agreement. That will the focus of this paper, provide the implementation and support to the level best that is without flaw (Target <1%). That can be attained when the process is flowed as mentioned in the paper.

## REFERENCES

The Shortcut Guide to Improving IT Service Support through ITIL

http://nexus.realtimepublishers.com/sgitil.php

ITIL and Value Network Analysis Laurence Lock Lee, Optimice Pty Ltd,

http://www.openvaluenetworks.com/Articles/VNA%20and%20ITIL.pdf

Specifying Services for ITIL Service Management

Alain Wegmann1, Gil Regev2, Georges-Antoine Garret3, François Maréchal4 1, 2 Ecole

Polytechnique Fédérale de Lausanne (EPFL),School of Communication and Computer

Science, CH-1015 Lausanne, Switzerland 3 Itecor, Avenue Paul Cérésole 24, CH-1800

Vevey, Switzerland4 SIG, Geneva, Switzerland

ITIL V3 Improves Information Security Management Ginger TaylorEast Carolina University

http://www.infosecwriters.com/text_resources/pdf/GTaylor_ITIL.pdf

Guidance on Aligning COBIT, ITIL and ISO 17799

http://www.isaca.org/Journal/Past-Issues/2006/Volume-1/Documents/jpdf0601-Guidance-on-

Aligning.pdf